\documentclass[twocolumn,seceq]{jpsj2}
\def\runtitle{Deviations from Berry--Robnik Distribution Caused by Spectral Accumulation}
\def\runauthor{Hironori \textsc{Makino} and Shuichi \textsc{Tasaki}}
\title
{Deviations from Berry--Robnik Distribution
Caused by Spectral Accumulation}
\author
{ Hironori \textsc{Makino}$^{1}$ and Shuichi \textsc{Tasaki}$^2$}

\inst {$^{1}$Department of Human and Information Science, Tokai
University, C-216, 1117 Kitakaname, Hiratsuka-shi, Kanagawa 259-1292, Japan\\
$^{2}$Department of Applied Physics, Waseda University, 3-4-1
Okubo, Shinjuku-ku, Tokyo 169-8555, Japan
}
\recdate{January 31, 2003; Revised March 20, 2003; Revised April 12, 2003; Accepted April 14, 2003}
\abst {By extending the Berry--Robnik approach for the nearly integrable quantum systems,\cite{[1]} we propose one possible scenario of the energy level spacing distribution that deviates from the Berry--Robnik distribution.  The result described in this paper implies that deviations from the Berry--Robnik distribution would arise when energy level components show strong accumulation, and otherwise, the level spacing distribution agrees with the Berry--Robnik distribution.}
\kword {energy level statistics, Berry--Robnik distribution, infinitely many independent components }
\begin{document}
\sloppy
\maketitle
\section{Introduction}
For bounded quantum systems with two degrees of freedom, the statistical properties of the eigenenergy levels at a given small energy interval have been intensively studied.
Universal behaviors have been found in the statistics of {\it unfolded} energy levels, which are the sequence of numbers uniquely determined by 
the energy levels using the mean level density obtained from the Thomas-Fermi rule.\cite{[2]}  For quantum systems whose classical counterparts 
are integrable, the distribution of nearest-neighbor level spacing is characterized by the Poisson distribution,\cite{[3],[4]} while for quantum systems 
whose classical counterparts are strongly chaotic, the level statistics are well characterized by the random matrix theory which gives level-spacing distribution 
obeying the Wigner distribution.\cite{[5],[6]}  The generic dynamical system is neither integrable nor strongly chaotic whose phase space consists 
of tori and chaos.  In this paper, such a system will be referred to as a nearly integrable system.  For several nearly integrable systems, the level 
spacing distribution of quantum systems is well characterized by the Berry--Robnik distribution.\cite{[7],[71],[72],[73],[74],[75]}

The Berry--Robnik distribution is derived by assuming that the energy level spectrum is a product of the superposition of independent subspectra, which are 
contributed respectively from localized eigenfunctions onto invariant (disjoint) phase space regions.  The formation of such independent subspectra is expected to be justified by {\it{the principle of uniform semi-classical condensation of eigenstates}},\cite{Li,[8]} which is based on an implicit state by Berry\cite{[9]}.

This principle states that the Wigner function of a semiclassical eigenstate is connected to a region in phase space explored by a typical trajectory of the classical dynamical system.  In the integrable system, the phase space is foliated into invariant tori, and the Wigner functions tend to delta functions on these tori in the semiclassical limit.\cite{[10]}  On the other hand, in a strongly chaotic system, almost all trajectories cover the energy shell uniformly, and hence the Wigner functions are expected to be a delta function on the energy shell, as suggested by the quantum ergodicity theorem.\cite{[11],[12]}  The nearly integrable system contains both properties of these two systems, and in the semiclassical limit, because of the suppression of tunneling, the Wigner function localizes each one of tori and chaotic regions in the phase space.

The Basic idea proposed by Berry and Robnik is to connect the compartmentalization of each eigenfunction with the independence between energy level components, which is supposed to be justified by lack of mutual overlap between eigenfunctions localizing onto different phase space regions. Therefore, in the Berry--Robnik theory, the semiclassical limit is one of the mechanisms providing independent spectral components.

On the basis of the above arguments, Berry and Robnik derived the level spacing distribution for the nearly integrable quantum system as follows.

Consider a system whose classical phase space is decomposed into $N$ disjoint regions.  The Liouville measures of these regions are denoted by
$\rho_i\,(i=1,2,3,\ldots,N)$ which satisfy $\sum_{i=1}^N\rho_i= 1$.  Let $E(S)$ be the gap distribution which stands for the probability that an interval $(0,S)$ contains no level.\cite{[6]}  $E(S)$ is expressed by the level spacing distribution as
%
$$
E(S)=\int_S^{+\infty} d\sigma \int_{\sigma}^{+\infty} P(x)\,dx.
$$
%
When the entire sequence of energy levels is a product of the
statistically independent superposition of $N$ sub-sequences,
$E(S;N)$ is decomposed into those of
sub-sequences, $E_i(S;\rho_i)$,
%
\begin{equation}
E(S;N)=\prod_{i=1}^N E_i(S;\rho_i).
\label{eq:1-2}
\end{equation}
%
In terms of the normalized level
spacing distribution $p_i(S;\rho_i)$ of
a sub-sequence, $E_i(S;\rho_i)$ is given by
$$
E_i(S;\rho_i)= \rho_i\int_S^{+\infty} d\sigma
\int_\sigma^{+\infty} p_i(x;\rho_i)\,dx,
$$
and $p_i(S;\rho_i)$ is assumed to satisfy\cite{[1]}
%
\begin{equation}
\int_0^{+\infty} S\cdot p_i(S;\rho_i)\,dS =\frac{1}{\rho_i}.
\label{eq:1-4}
\end{equation}
%
Equation ({\ref{eq:1-4}}) is satisfactory when the Thomas-Fermi rule for individual phase space regions still holds.\cite{[1],[MK1]}  eqs.~({\ref{eq:1-2}}) and ({\ref{eq:1-4}}) relate the level statistics in the semiclassical limit with the phase-space geometry, and provide a generic form of the level spacing distribution.

In the most general cases, the level spacing distribution might be singular.  In such a case, it is convenient to use its cumulative distribution function $\mu_i$;
\begin{equation}
\mu_i(S)=\int_0^S p_i(x;\rho_i)\,dx.
\label{eq:1-5}
\end{equation}
The corresponding quantity of the overall spectrum is
\begin{equation}
M(S;N)= \int_0^S P(x;N)\,dx
= 1+\frac{d}{dS}E(S;N).
\label{eq:1-6}
\end{equation}
where $P(S;N)$ is the level spacing distribution function corresponding to $E(S;N)$.

Here, let us consider the phase space geometry of the generic nearly integrable systems.

The phase space of the nearly integrable system consists of infinitely many sub-domains in the regular region and a finite number of chaotic regions which are disconnected by tori, respectively.  Hereafter we rewrite the Liouville measure of each sub-domain in the regular region by $\rho_i\,(i=1,2,3,\ldots,N_1)$ and that of each chaotic region by $\rho_j'\,(j=1,2,3,\ldots,N_2)$ which satisfies $\sum_{i=1}^{N_1}\rho_i\equiv \rho_0$, and $\rho_0 +\sum_{j=1}^{N_2}\rho_j'=1$.  For each element of the spectral components, Berry and Robnik introduced the following two surmises\cite{[1]}:
\begin{itemize}
\item
Berry--Robnik surmise (i): The level-spacing distributions of the spectral components contributed from chaotic regions are characterized by the random matrix theory.  We denote the gap distribution function obeying the random matrix theory by
$E_j^{\mbox{{\tiny{RMT}}}}(S;\rho_j')$.
\item
Berry--Robnik surmise (ii): The level-spacing distribution of the spectral component contributed from the regular region is characterized by the Poisson distribution.
\end{itemize}

As suggested, e.g., by Hannay (see Sec.~4 of ref.~1), the Berry--Robnik surmise (ii) would be interpreted as follows: In the regular phase space region, almost every orbit is generically confined in each inherent torus, and the whole regular region is densely covered by invariant tori.  The spectral component is then a superposition of infinitely many subsequences which are contributed from those regions.
Therefore, if the mean level spacing of each independent subset is large, one would expect the Poisson distribution to be a result of the law of small numbers.\cite{[13]}  Along the lines of this scenario, the Berry--Robnik surmise (ii) is described by the following limit:
%
%
$
\lim_{N_1\to+\infty}\prod_{i=1}^{N_1}E_i(S;\rho_i)
= e^{-\rho_0 S}.
$

%
%

Therefore, by applying the Berry--Robnik surmises (i) and (ii) to eq.~({\ref{eq:1-2}}), the overall level spacing distribution reduces to the Berry--Robnik distribution whose cumulative distribution is given by the following function:
\begin{equation}
M^{\mbox{\tiny BR}}(S;N_2)=1+\frac{d}{dS}\left[
e^{-\rho_0 S}\prod_{j=1}^{N_2}
E_j^{\mbox{\tiny{RMT}}}(S;\rho_j')\right].
\label{M_BR}
\end{equation}

In this paper, by taking into account of the infinitely many infinitesimal structures in the regular region, the Berry--Robnik surmise (ii) is extended along the line of thought of Hannay's proposal.  Instead of beginning the Berry--Robnik surmise (ii), we start with the following two assumptions:
\begin{itemize}
\item Assumption (1): The statistical weights
of regular regions uniformly vanish in the limit
of infinitely many regions;
\begin{equation}
\max_i \rho_i \to 0\quad\mbox{as}\quad N_1\to +\infty.
\label{eq:1-7}
\end{equation}
\item Assumption (2): The weighted mean of
the cumulative distribution of energy spacing,
\begin{equation}
\mu(\rho;N_1)=\sum_{i=1}^{N_1}\rho_i\mu_i(\rho),
\label{eq:1-8}
\end{equation}
converges in $N_1\to +\infty$ to $\bar{\mu}(\rho)$
\begin{equation}
\lim_{N_1\to +\infty}\mu(\rho;N_1) = \bar{\mu}(\rho).
\label{eq:1-9}
\end{equation}
The limit is uniform on each closed interval: \ $0\le \rho \le S$.
\end{itemize}

Under Assumptions (1) and (2) and the Berry--Robnik surmise(i),
eqs.~({\ref{eq:1-2}}) and ({\ref{eq:1-4}}) lead to the overall level spacing distribution whose cumulative distribution function is given by the following formula in the limit of $N_1\to +\infty$,
%
\begin{align}
M_{\bar{\mu}}(S;N_2)=
1+\frac{d}{dS}\,\Biggl[&
\exp\biggl(
-\rho_0\int_0^S[1-\bar{\mu}(\sigma)]\,d\sigma
\biggr)\nonumber\\
&{}\times\prod_{j=1}^{N_2} E_j^{\mbox{\tiny{RMT}}}(S;\rho_j')
\Biggr],
\label{eq:1-10}
\end{align}
%
where the convergence is in the sense of the weak limit.

When the level spacing distributions
of individual components contributed from the regular
regions are sparse enough, one may expect $\bar{\mu}=0$
and the level spacing distribution of the whole energy sequence
is reduced to the Berry--Robnik distribution.  In general, one may expect $\bar{\mu}\not=0$ which corresponds to a certain accumulation of the levels of individual components.

In the following sections, the above statement is proved and the limiting level spacing distributions are classified into three classes.  One of them is the Berry--Robnik distribution.\cite{[1]}  The others are not the Berry--Robnik
distribution.  We give a condition in which the distribution formula
({\ref{eq:1-10}}) agrees with the Berry--Robnik distribution, and also
discuss a situation in which the limiting
distribution formula ({\ref{eq:1-10}}) shows deviations from the
Berry--Robnik distribution.
\section{Limiting level spacing distribution}
\subsection{Derivation of the limiting level spacing distribution}
In this section, starting from eqs.~({\ref{eq:1-2}}) and ({\ref{eq:1-4}}),
the Berry--Robnik surmise (i), and the additional Assumptions (1)
and (2) introduced in the previous section, we show that, in the
limit of infinitely many components $N_1\to +\infty$ $(N_2\ll+\infty)$, the level-spacing distribution converges weakly
to the distribution with the cumulative distribution function
(\ref{eq:1-10}).  According to Helly's theorem,\cite{[13]}
this is equivalent to show the limit:
$\lim_{N_1\to+\infty}M(S;N_1,N_2)=M_{\bar{\mu}}(S;N_2)$.
The convergence is shown as follows.

With the aid of eqs.~({\ref{eq:1-4}}) and (\ref{eq:1-5}),
we rewrite $E_i(S)$ in terms of $\mu_i(S)$:
$$
E_i(S)=\rho_i \int_S^{+\infty} d\sigma\,[1-\mu_i(\sigma)].
$$
The overall cumulative level spacing distribution function $M(S;N_1,N_2)$ is then given by
\begin{align}
&M(S;N_1,N_2)\nonumber\\
&\ =1+ {d\over dS}E(S;N_1,N_2)\nonumber\\
&\ =1+ \Biggl[
-\sum_{i=1}^{N_1}
\frac{1-\mu_i(S)}{\int_S^{+\infty} d\sigma\,
[1-\mu_i(\sigma)]}\prod_{j=1}^{N_2}
E_j^{\mbox{\tiny{RMT}}}(S)\nonumber\\
&\ \phantom{=1+ \Biggl[}
{}+\frac{d}{dS}\prod_{j=1}^{N_2}E_j^{\mbox{\tiny{RMT}}}(S)\Biggr]
\prod_{i=1}^{N_1} \rho_i\int_S^{+\infty}d\sigma\,[1-\mu_i(\sigma)].
\label{eq:2-1}
\end{align}
First we consider the behavior of
$\prod_{i=1}^{N_1} \rho_i\int_S^{+\infty}d\sigma[1-\mu_i(\sigma)]$.  >From eq.~(1.4), integration by
parts, and $\lim_{\sigma\to+\infty}\sigma[1-\mu_i(\sigma)]=0$
which follows from the existence of the average, one has
$$
\rho_i\int_S^{+\infty}d\sigma\,[1-\mu_i(\sigma)]=
1-\rho_i\int_0^S d\sigma\,[1-\mu_i(\sigma)].
$$

Since the convergence of $\sum_{i=1}^{N_1} \rho_i
\mu_i(S) \to \rho_0{\bar \mu}(S)$ for $N_1\to +\infty$
is uniform by Assumption (2),
\begin{align}
&\log{\prod_{i=1}^{N_1} \rho_i\int_S^{+\infty}d
\sigma\,[1-\mu_i(\sigma)]}\nonumber\\
&\quad=\sum_{i=1}^{N_1}\log{\left[1-\rho_i\int_0^{S}d\sigma\,
[1-\mu_i(\sigma)]\right]}\nonumber\\
&\quad=-\sum_{i=1}^{N_1}\left[\rho_i\int_0^{S}d\sigma\,
[1-\mu_i(\sigma)]+O(\rho_i^2)\right]\nonumber\\
&\quad=-\rho_0\int_0^S d\sigma\,
[1-\mu(\sigma;N_1)]
+\sum_i^{N_1} O(\rho_i^2)\nonumber\\
&\quad\longrightarrow -\rho_0\int_0^S d\sigma\,
[1-\bar{\mu}(\sigma)]
\quad\mbox{as}\quad N_1\to+\infty
\label{lim:01},
\end{align}
where we have used $|\mu_i(\sigma)|\le 1$, $\log(1+\epsilon)=\epsilon+O(\epsilon^2)$ in $\epsilon\ll 1$ and the following property obtained from Assumption (1),
\begin{align}
\left|\sum_{i=1}^{N_1} O(\rho_i^2)\right|
&\leq C\cdot\max_i{\rho_i}\cdot\sum_{j=1}^{N_1}\rho_j \nonumber\\
&=C\rho_0\cdot\max_i{\rho_i}\to 0\quad\mbox{as}
\quad {N_1}\to+\infty\nonumber,
\end{align}
with $C$ a positive constant.

The $N_1\to +\infty$ limit of the sum in the
right-hand side of (\ref{eq:2-1})
can be calculated in a similar way. Indeed,
as $1/(1-\epsilon)=1+O(\epsilon)$ in $\epsilon\ll 1$, one has
\begin{align}
\sum_{i=1}^{N_1} {1-\mu_i(S)\over \int_S^{+\infty} d\sigma\,
[1-\mu_i(\sigma)]}
&= \sum_{i=1}^{N_1} {\rho_i-\rho_i \mu_i(S)\over
1-\rho_i\int_0^{S}d\sigma\,[1-\mu_i(\sigma)]}\nonumber\\
&=\rho_0 -\sum_{i=1}^{N_1}\rho_i\mu_i(S)
+\sum_{i=1}^{N_1} O(\rho_i^2)\nonumber\\
&\longrightarrow \rho_0[1-\bar{\mu}(S)]
\ \mbox{as}\ N_1\to+\infty.\nonumber
\end{align}
Therefore, we have
$M(S;N_1,N_2)\to M_{\bar{\mu}}(S;N_2)$ as $N_1\to+\infty$.
%
%
%
%
%
%
\subsection{Property of the limiting level spacing distribution}
Since $\mu_i(S)$ increases monotonically
and $0 \le \mu_i(S) \le 1$,
$\bar{\mu}(S)$ has the same properties.
Then, $1-\bar{\mu}(S)\ge 0$ for any
$S\ge 0$ and one has
\begin{equation}
\frac{1}{S}\int_0^S d\sigma\,[1-\bar{\mu}(\sigma)]\longrightarrow 1-\bar{\mu}(+\infty)
\quad\mbox{as}\quad S\to+\infty.
\label{eq:4-7}
\end{equation}
%
%
%
%
According to the above limit, the level spacing distribution is classified into the following three cases in the sense of weak limit:
\begin{itemize}
\item Case 1,
$\bar{\mu}(+\infty)=0$: The limiting level spacing distribution is the
Berry--Robnik distribution.  Note that this condition is equivalent to
$\bar{\mu}(S)=0$ ${}^{\forall} S$ because $\bar{\mu}(S)$ increases monotonically.
\item Case 2, $0<\bar{\mu}(+\infty)< 1$:
For large $S$ values, the limiting level spacing distribution is well approximated by the Berry--Robnik distribution, but possibly not for small $S$ values.
\item Case 3,
$\bar{\mu}(+\infty)=1$:
The limiting level spacing distribution is a product of the superposition of spectral components obeying the sub-Poisson statistics and the Random matrix theory, and deviates from the Berry--Robnik distribution ${}^{\forall} S$.
\end{itemize}
One has Case 1 if the individual level spacing distributions are derived from scaled distribution functions $f_i$ as
\begin{equation}
\mu_i(S) =\rho_i\int_0^S
f_i\left(\rho_i x\right)dx ,
\label{eq:4-8}
\end{equation}
where $f_i$ satisfy
$$
\int_0^{+\infty} f_i(x) \,dx = 1, \quad\int_0^{+\infty} x f_i(x) \,dx = 1,
$$
and are uniformly bounded by a positive constant $D$: $|f_i(S)|\le D$
($1\le i \le N_1$ and $S\ge 0$).  Indeed, one then has
\begin{align}
|\mu(S;N_1)|
&\leq \sum_{i=1}^{N_1} \rho_i^2\int_0^S
\left| f_i\left({\rho_i}x\right)\right|dx \nonumber \\
&\leq DS\max_i\rho_i\sum_{j=1}^{N_1}\rho_j
\longrightarrow 0\equiv {\bar \mu}(S).
\nonumber
\end{align}
In general, one can expect $\bar{\mu}(S)\not=0$ which corresponds
to deviations from the Berry--Robnik distribution.  Such a case is expected when there is strong accumulation of the energy levels of individual components which leads to the non-smooth cumulative distribution function $\mu_i(S)$.

We remark that, when the limiting function
${\bar \mu}(S)$ is differentiable, the asymptotic level spacing distribution can be described as
\[
P_{\bar{\mu}}(S)= \frac{d^2}{dS^2}\Biggl[
\exp\biggl(
-\rho_0\int_0^S[1-\bar{\mu}(\sigma)]\,d\sigma
\biggr)\\
\prod_{j=1}^{N_2}E_j^{\mbox{\tiny{RMT}}}(S)
\Biggr].
\]
\section{Summary}
In this paper, we have proposed one possible scenario that the level spacing distribution of the nearly integrable quantum system deviates from the Berry--Robnik distribution.  By extending the Berry--Robnik theory, the limiting level spacing distribution for the nearly integrable quantum systems was obtained which is described by a single monotonically increasing function $\bar{\mu}(S)$ of the level spacing $S$.  The limiting distribution is classified into three cases:
Case 1: Berry--Robnik distribution if $\bar{\mu}(+\infty)=0$,
Case 2: Berry--Robnik distribution for large $S$, but possibly not for small $S$ if $0<\bar{\mu}(+\infty)< 1$, and Case 3: the level spacing distribution given by the superposition of spectral components obeying the sub-Poisson statistics and the random matrix theory if $\bar{\mu}(+\infty)=1$.  Thus, even when the energy levels of individual components are statistically independent, deviations from the Berry--Robnik distribution are possible.  The result described in this paper implies that the deviations would arise when the spectral component contributed from the regular phase space regions shows the strong accumulation which leads to the non-smooth cumulative level spacing distribution, and otherwise, the level spacing distribution agrees with the Berry--Robnik distribution.

For several nearly integrable quantum systems, the level spacing distribution is investigated numerically.\cite{[7],[71],[72],[73],[74],[75]}
Among them, Prosen and Robnik found that there is a high energy region in which the Berry--Robnik distribution formula well approximates the level spacing distribution of the nearly integrable quantum system.  This energy region is sometimes referred to as {\it the Berry--Robnik regime}.\cite{[72]}  While they also found that the level spacing distribution in the low energy region deviates from the Berry--Robnik distribution, and approximates the Brody distribution.  This behavior was studied numerically in terms of a fractional power dependence of the spacing distribution near the origin at $S=0$, which could be attributed to the localization properties of eigenstates on chaotic components.\cite{[7],[71]}
>From our results in this paper, Case 1: $\bar{\mu}(+\infty)=0$ should be satisfied in the Berry--Robnik regime.  While Case 2 and Case 3 might propose another possibilities.  When the spectral components corresponding to the regular regions show strong accumulation, the statistical property of energy levels obeys the cumulative distribution formula (1.9) with $0<\bar{\mu}(+\infty)\leq 1$, and the level spacing distribution shows deviations from the Berry--Robnik distribution.  Therefore, the results shown in this paper might propose another possibilities of the Berry--Robnik approach.

The accumulation of spectral components would be enhanced when the system has an inherent symmetry.  For several integrable quantum systems ($\rho_0=1$), such accumulation is possible.  One known example is the rectangular billiard.\cite{[14],[MK1],[MK2]}   The level spacing distribution of this system deviates from the Poisson distribution when the aspect ratio of two sides of a billiard wall is close to a rational.  Another example is studied by Shnirelman, Chirikov and Shepelyansky, and Frahm and Shepelyansky for a certain type of system which contains a quasi-degeneracy result from inherent symmetry (time reversibility).\cite{[15],[16],[17]}  As is well known, the existence of quasi-degeneracy leads to the sharp peak at small level spacings.  These phenomena might arise also in the spectral component of the nearly integrable quantum systems.
Such possibilities in the physical systems will be studied elsewhere.
\section*{Acknowledgments}
The authors would like to thank Professors M. Robnik, Y. Aizawa and A. Shudo for their helpful advice.
The authors also thank the Yukawa Institute for Theoretical
Physics at Kyoto University.  Discussions
during the YITP workshop YITP-W-02-13 on ``Quantum chaos:
Present status of theory and experiment'' were useful to
complete this work.   This work is partly supported by
a Grant-in-Aid for Scientific Research
(C) from the Japan Society for the
Promotion of Science.

\end{document}